# Structural and Optical Characteristics of $\beta$-Ga$_2$O$_3$ Implanted with Rare Earth Ions

*Renata Ratajczak[1]\*, Joanna Matulewicz[1,4], Slawomir Prucnal[2], Maciej O. Liedke[2], Cyprian Mieszczynski[1], Przemyslaw Jozwik[1], Ulrich Kentsch[2], Rene Heller[2], Eric Hirschmann[2], Andreas Wagner[2], Wojciech Wozniak[3], Frederico Garrido[4] and Elzbieta Guziewicz[3]*

[1] *National Centre for Nuclear Research, Andrzeja Soltana 7, 05-400 Otwock-Swierk, Poland*
[2] *Helmholtz-Zentrum Dresden-Rossendorf, Bautzner Landstrasse 400, D-01328 Dresden, Germany*
[3] *Institute of Physics, Polish Academy of Sciences, Aleja Lotnikow 32/46, 02-668 Warsaw, Poland*
[4] *Universite Paris-Saclay, CNRS/IN2P3, IJCLab, 91405 Orsay, France*

**Abstract**

We investigated the structural evolution and optical properties of $\beta$-Ga$_2$O$_3$ crystals implanted with different rare earth (RE) ions using channeling Rutherford Backscattering Spectrometry, Positron Annihilation, Photoluminescence, and Photoluminescence Excitation Spectroscopies. The studies presented here reveal that implantation-induced disorder, accompanying phase transitions, and post-annealing structural recovery are insensitive to the implanted RE species. The defect microstructure also remains similar for all RE ions. However, thermal annealing does not completely remove radiation-induced defects but instead drives their rearrangement into larger defect complexes. It is shown that unimplanted (virgin) $\beta$-Ga$_2$O$_3$ exhibits strong UV–visible emission attributed to oxygen vacancies, while the introduction of RE ions gives rise to additional emission lines, originating from the transition of a given RE$^{3+}$ ion. Moreover, our results suggest that all RE$^{3+}$ are excited via the host conduction band, followed by non-radiative relaxation to the 4f excited states and radiative decay to the respective ground states. In turn, fluences-dependent studies of Yb$^{3+}$ reveal the onset of concentration quenching, and RE-related emission remains efficient despite substantial lattice disorder. The results provide new insight into defect evolution in ion-implanted β-Ga$_2$O$_3$ and the excitation mechanisms of RE$^{3+}$ ions, offering valuable guidance for optimizing the optical performance of β-Ga$_2$O$_3$:RE systems.

**1 Introduction**

β-Ga$_2$O$_3$ is a wide-bandgap oxide semiconductor with exceptional thermal, chemical, and radiation resistance, making it a prime candidate for deep-UV photonic and power electronic applications in nuclear power plants and space systems. The published values of β-Ga$_2$O$_3$ band gap vary between 4.4 and 5.1 eV, depending on the fabrication process of the material and the specific method employed for band gap estimation [1–5]. These band gap values correspond to emission within the ultraviolet (UV) spectral region. However, the as-grown single-crystalline oxide is typically n-type due to cross-contamination and some defects. The most common cross-contaminations are Si and Sn, which are shallow donors, while the most common defect is oxygen vacancy [6, 7]. In addition, the existence of the self-trapped holes' energy level makes the p-type doping of β-Ga$_2$O$_3$ very challenging [8, 9]. Both defects and non-intentional cross-contaminations have a strong influence on the optical and electrical properties of the semiconductor.

Intentional incorporation of transition metals such as Mn or Cr [10] and rare-earth (RE) ions [11] offers a promising route for engineering the optical properties of β-Ga$_2$O$_3$, enabling the tuning of its intrinsic emission toward the visible and near-infrared spectral regions. The emission spectrum depends on the oxidation state of metals Mn$^{2+/3+}$ and Cr$^{3+}$, resulting in strong red emission. Most of RE in oxide matrix exists in 3+ oxidation state with 4f electrons screened from the host crystal field by outer 5s and 5p electrons. Such an electronic configuration of RE$^{3+}$ ions in oxides is responsible for strong and sharp photoluminescence with the optical emission ranging from UV to near-infrared, and the relative independence of their optical properties from the host matrix [5, 12-17]. However, wide bandgap semiconductors are well-suited to host RE ions due to efficient suppression of the thermal quenching and thus enhancing RE luminescence efficiency [18].

Due to the low solubility limit, the in-situ incorporation of RE dopant into the oxide matrix is challenging as it often leads to segregation or the formation of secondary phases [19-20]. Recent advances in β-Ga$_2$O$_3$ research have revealed that ion implantation is a promising method for incorporating dopants [21]. The controlled nature of ion implantation enables precise tuning of the dopant concentration and in-depth distribution. However, ion implantation inherently introduces lattice disorder, necessitating thermal treatment for structural recovery and optical activation of the incorporated RE ions [22].

Many experimental and theoretical works on damage accumulation in ion-bombarded β-Ga$_2$O$_3$ indicate a radiation-induced phase transition from β- to γ-phase at around 0.25 displacements per atom (dpa) [24-31]. These observations are in line with the studies by Matulewicz et al. on (-201)-oriented β-Ga$_2$O$_3$ implanted with Yb [32]. The authors have already obtained a nearly complete picture of the damage behavior in such systems using combined Rutherford Backscattering Spectrometry in channeling mode (RBS/c), Transmission Electron Microscopy (TEM), and X-Ray Diffraction (XRD). It was observed that for Yb-implanted β-Ga$_2$O$_3$ with a fluence of $1\times10^{14}$ ions/cm² (~0.7 dpa), the entire implanted layer transformed into the γ-phase. Such radiation-induced phase transformations are driven by lattice strain and seem to be independent of the crystal orientation [25, 32]. However, in the case of Yb-implanted β-Ga$_2$O$_3$ crystals, near-surface disorder has been observed under further irradiation, which subsequently evolves into a second phase transformation from the γ phase to an amorphous structure at approximately 7 dpa [32-33]. Amorphization as the reason for elevated backscattering yield at the random level in Eu-implanted β-Ga$_2$O$_3$ with $1\times10^{15}$ ions/cm² (~7 dpa) has already been suggested by Peres et al. [16] and is also likely responsible for the disappearance of the γ-phase signal detected by XRD after boron ion implantation to a fluence of about 7 dpa, as recently observed by Nikolskaya et al. [31]. In contrast, other authors have suggested that the γ-phase exhibits high radiation tolerance, remaining stable up to doses of approximately 260 dpa [28, 29]. These imply that the stability of the γ-phase may depend on irradiation conditions, implanted ions, or both.

It has been found that both of these radiation-induced phases are thermally unstable and disappear after annealing at approximately 800 °C [22, 30, 33]. Although in the case of Yb-implanted β-Ga$_2$O$_2$, annealing at 800 °C restores the β-phase and eliminates the radiation-induced γ and amorphous phases, the recovery is not complete, and the regrown structure suggests the existence of void/cluster-like defects in the regrown β-Ga$_2$O$_3$ layer as well as misorientation compared to the pristine β-Ga$_2$O$_3$ lattice [30]. Post-annealing luminescence studies reveal that RE ions retain high luminescence efficiency despite considerable lattice disorder [5, 15-16]. The optimal annealing conditions ensuring the highest RE luminescence were established using a rapid thermal annealing (RTA) process at 800 °C for 10 min in an oxygen atmosphere [22].

As it was mentioned, the origin of RE-related photoluminescence is relatively well understood. However, the excitation mechanisms responsible for energy transfer between the host lattice and the RE ions remain a subject of ongoing debate. Two non-consistent models have been proposed to explain the mechanism of intra-4f $RE^{3+}$ luminescence in $\beta$-$Ga_2O_3$. Recent studies by Zanoni et al. on Pr-doped gallium oxide and by Tokida and Adachi on $Tb^{3+}$ in $\beta$-$Ga_2O_3$ suggest that the 5d energy level is involved in the excitation mechanism [5, 15]. It has been postulated that for $Tb^{3+}$ and $Pr^{3+}$ ions in $\beta$-$Ga_2O_3$, electrons are first excited to the 5d shell located within the conduction band of gallium oxide, followed by non-radiative relaxation from the 5d to the 4f excited states, and finally radiative transitions to the $4f^2$ ground state [5]. In contrast, the excitation mechanism proposed for $Tm^{3+}$ ions by Zewei Chen et al. involves a predominantly defect-assisted energy transfer process from the $\beta$-$Ga_2O_3$ host lattice to the $Tm^{3+}$ 4f states [35].

In this context, the present study extends these findings by systematically investigating the effects of dysprosium (Dy), erbium (Er), and yttrium (Yb) ion implantation on defect formation, structural recovery dynamics, and the excitation mechanisms governing the optical emission of $RE^{3+}$ ions in $\beta$-$Ga_2O_3$ films. It was found that different RE ions, despite similar masses, interact differently with the ZnO lattice during annealing [13]. Consequently, they were selected to ascertain whether this phenomenon also applies to $\beta$-$Ga_2O_3$. The samples were characterized using Rutherford Backscattering Spectrometry in channeling mode (RBS/c), Doppler Broadening Variable Energy Positron Annihilation and Variable Energy Positron Annihilation Lifetime Spectroscopies (DB-VEPAS and VEPALS), and Photoluminescence (PL) as well as Photoluminescence Excitation (PLE) techniques. Based on correlating the microstructural evolution with the luminescence response, this work provides new insights into the RE-activation pathways and the interplay between native and implantation-induced defects, as well as the optical performance of $\beta$-$Ga_2O_3$:RE systems. The results demonstrate that the structural modifications observed after implantation and subsequent annealing are largely independent of the implanted ion species. Moreover, our results suggest the optical excitation mechanism is also the same for all RE in 3+ oxidation state embedded into $\beta$-$Ga_2O_3$, and it is not necessarily the non-radiative relaxation from the 4f5d state to the 4f levels as recently proposed [5]. It is also largely independent of implantation-induced structural defects and subsequent annealing, indicating that radiative processes are primarily controlled by direct RE–host interactions rather than by defect-mediated transitions.

## 2 Experimental Section

The (-201) oriented β-Ga$_2$O$_3$ commercial n-type single crystal with an electron concentration below 9 × 10$^{17}$ cm$^{−3}$ manufactured by Tamura Corporation (Japan) was used in the present studies. The purchased 2-inch, 0.68 mm-thick wafer was diced into smaller samples measuring 9 × 9 mm. Subsequently, the samples were implanted at the Ion Beam Center (IBC), Helmholtz-Zentrum Dresden-Rossendorf (HZDR), Germany with three different species of RE$^+$ ions, such as dysprosium (Dy), erbium (Er), and yttrium (Yb) with an energy of 150 keV and with fluences of 1 × 10$^{13}$, 1 × 10$^{14}$, and 1 × 10$^{15}$ cm$^{−2}$. These fluences were selected based on our comprehensive studies of damage accumulation in Yb-implanted β-Ga$_2$O$_3$, as they correspond to implantation fluencies that induce the structural phase transformation β- γ-AMO [27, 32].

Subsequently, the implanted crystals were subjected to rapid thermal annealing (RTA) in an O$_2$ atmosphere at 800 °C for 10 minutes using an AccuThermo AW-610 system (Allwin21 Corporation) at the Institute of Physics, Polish Academy of Sciences (IP PAS). These annealing conditions were found to enable only partial lattice recovery, but provide the most intense RE-related emission in β-Ga$_2$O$_3$ [22].

Rutherford Backscattering Spectrometry in channeling mode (RBS/c), Doppler broadening variable energy positron annihilation spectroscopy (DB-VEPAS), and variable energy positron annihilation lifetime spectroscopy (VEPALS) were employed to analyze implantation-induced damage and subsequent structural recovery after annealing, while photoluminescence (PL) spectroscopy with λ$_{exc}$ ≈ 240 nm and photoluminescence excitation (PLE) technique was used to study the optical emission features of the implanted ions of the annealed samples. The RBS/c experiments were conducted at the IBC, HZDR, using a 2 MV Van de Graaff accelerator equipped with a silicon detector positioned at a 170° scattering angle, providing a depth resolution better than 5 nm and an energy resolution below 20 keV. DB-VEPAS measurements were carried out at the Institute of Radiation Physics, HZDR, using the Apparatus for In-situ Defect Analysis (AIDA) [48] integrated into the slow-positron beamline SPONSOR [49]. Positrons were accelerated and implanted monoenergetically into the samples over an energy range of Ep = 0.05–35 keV, enabling depth-resolved defect profiling. In turn, VEPALS measurements were conducted at the Mono-energetic Positron Source (MePS) beamline at HZDR [48]. A CeBr3 scintillator coupled to a Hamamatsu R13089-100 photomultiplier tube (PMT) was utilized for gamma photon detection. The signals were processed by the Teledyne SP Devices ADQ14DC-2X digitizer (14-bit vertical resolution and 2GS/s horizontal resolution) [50]. The overall time resolution of the measurement system was ≈0.230 ns, and all spectra contained at least $1 \times 10^7$ counts. Additional methodological details related to DB-VEPAS and VEPALS— including the procedures used to determine depth profiles, positron lifetimes, and other spectral parameters—are provided in Methods of the Supplementary Material. Finally, the optical properties of undoped and post-implanted annealed β-$Ga_2O_3$:RE samples were also studied at RT at HZDR. Samples were excited using monochromatic light from an Xe-lamp integrated with the spectrometer iHR 330, Horiba. The full width at half maximum of the PLE was about 6 nm. The PL signal was recoded using an iHR550 spectrometer and a photomultiplier Hamamatsu for the UV-visible spectral range (up to 900 nm) and an InGaAs detector for the near infrared range (800 – 1600 nm). In order to increase the signal-to-noise ratio, the lock-in amplifier was used. The excitation light was mechanically chopped at the frequency of 20 Hz. The PLE spectrum was collected in the range from 200 to 600 nm by monitoring the change in the intensity of the selected emission peak as a function of the wavelength of the excitation light. In order to reduce the influence of the second-order diffraction peaks of excitation light on the emission spectrum, the long-pass optical filters were used.

## 3 Results and Discussion

### 3.1 Defect build-up and structural recovery

A typical RBS/c spectrum obtained for β-$Ga_2O_3$ implanted with 150 keV RE ions consists of two characteristic regions: a low-energy part below 1380 keV, reflecting the scattering from host Ga atoms, and a high-energy peak in the 1510–1580 keV range, corresponding to the scattering from RE atoms (see Figure 1). The signal from the Ga substructure provides information about damage build-up and crystal structure recovery after annealing, whereas the high-energy peak reveals the depth distribution of RE atoms and their location in the host crystal lattice before and after annealing.

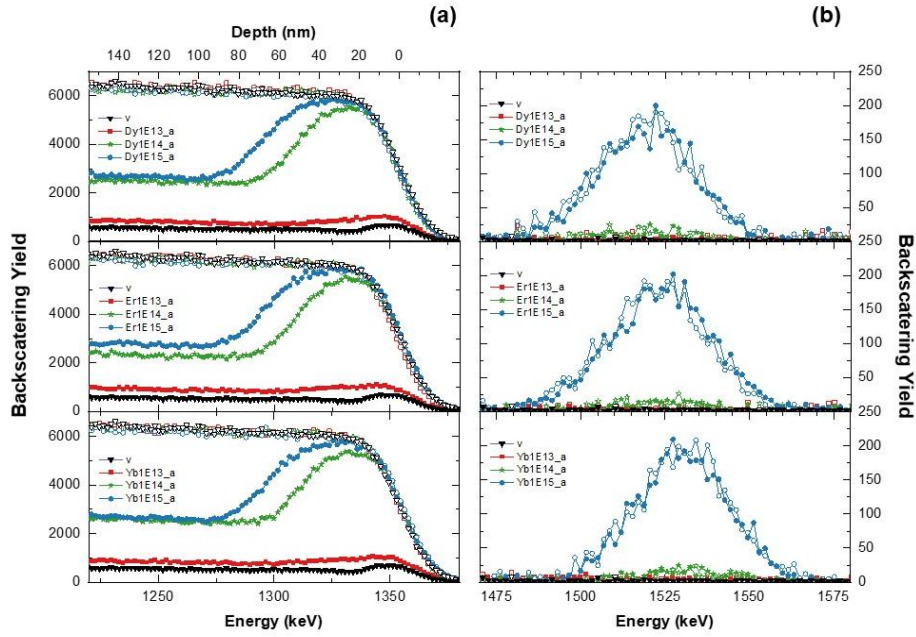

*Figure 1. Random (open symbols) and aligned (solid symbols) RBS experimental spectra obtained for (-201)-oriented β-Ga₂O₃ single crystals, virgin (unimplanted) and implanted with different fluences of Dy, Er, and Yb ions; (a) and (b) refer to signals coming from He⁺ backscattered on Ga atoms of the matrix and RE ions, respectively.*

The damage behavior in (-201)-oriented β-Ga$_2$O$_3$ implanted with different RE ions (Dy, Er, and Yb) to fluences of $1\times10^{13}$, $1\times10^{14}$, and $1\times10^{15}$ at/cm$^2$ is presented in Figure 1a. For comparison, the aligned spectrum of a virgin (unimplanted) sample is also shown. As can be seen, the characteristic damage peak observed between 1275 and 1340 keV increases in intensity with increasing fluence, indicating the growth of the disordering into the β-Ga$_2$O$_3$ crystals. For a fluence of $1\times10^{14}$ ions/cm$^2$, the damage almost reaches the random level, and for higher fluences, the damaged region becomes a double and extends deeper into the substrate with the clear saturation at the random level in the region near the surface (i.e., 0-30 nm). Such behavior is typical for Yb-implanted β-Ga$_2$O$_3$, as reported in previous studies [27, 32], where for a fluence of $1\times10^{14}$ ions/cm$^2$, the entire implanted layer was identified as γ-Ga$_2$O$_3$. The formation of γ-phase results from defect-induced structural transformation and has been observed by other researchers [28-31]. Further implantation with a fluence of $1\times10^{15}$ ions/cm² leads to a partial transformation of this γ-phase layer into an amorphous overlayer. By comparing the spectra obtained for different RE ions, it can be concluded that no significant differences in defect-formation behavior are observed among Dy, Er, and Yb implantation. Furthermore, looking at Figure 2 (also Figure 3, Figure S1, and Table 1), it could be concluded that there are no substantial differences in structural recovery observed after annealing, which is in contrast with the behavior of these RE ions in a ZnO matrix reported by Ratajczak et al. [13].

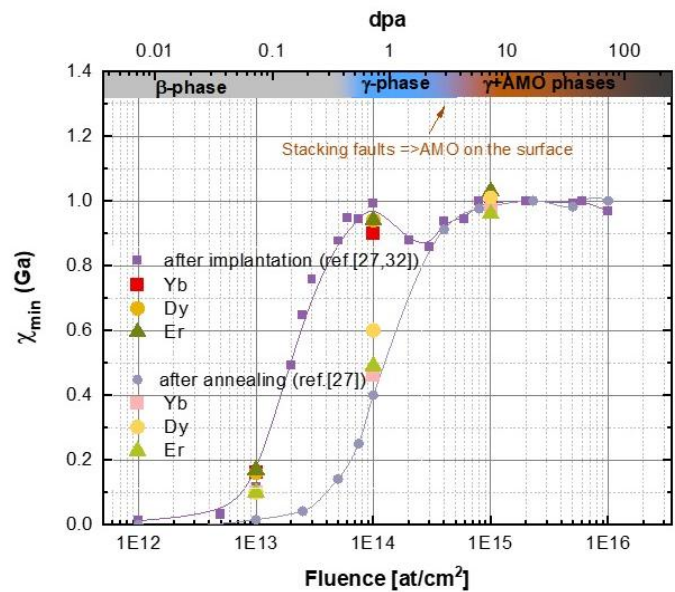

***Figure 2.*** *Comparison of defect accumulation behavior of Dy, Er, and Yb-implanted β-Ga$_2$O$_3$ prior to and after annealing.*

***Table 1*** *The calculated χ$_{min}$ values of the disorder level*

| Fluence (at/cm$^2$) | χ$_{min}$ (Ga) | | | | | | |
|---|---|---|---|---|---|---|---|
| | virgin | as implanted | | | annealed | | |
| | | Yb | Dy | Er | Yb | Dy | Er |
| 0 (virgin) | 0,08 | | | | | | |
| 1×10$^{13}$ | | 0,16 | 0,16 | 0,17 | 0,10 | 0,10 | 0,10 |
| 1×10$^{14}$ | | 0,90 | 0,94 | 0,94 | 0,46 | 0,60 | 0,49 |
| 1×10$^{15}$ | | 0,98 | 1,01 | 1,03 | 0,97 | 1,01 | 0,96 |

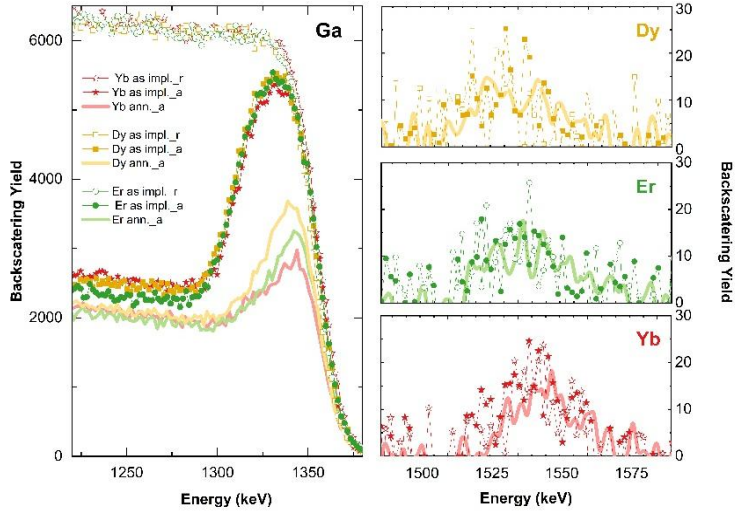

*Figure 3. Comparison of random and aligned RBS spectra obtained for the (-201)-oriented β-Ga$_2$O$_3$ single crystals implanted with Dy, Er, and Yb ions to a fluence of 1x10$^{14}$ /cm$^2$ prior to and after annealing in oxygen atmosphere at 800 °C for 10min.*

The x-scale in Figure 2 reflects the damage level observed in the aligned spectra of crystals implanted with different RE ions, normalized to the corresponding random level, referred to as the crystallographic quality parameter $\chi_{min}$ [23]. The results obtained here were compared with previous comprehensive studies on damage accumulation and annealing in (-201)-oriented β-Ga$_2$O$_3$ implanted with Yb [27, 32]. The calculated values are also summarized in Table 1. As shown, full structural recovery is achieved only for very low fluences, well below the threshold for the β→γ phase transition [26, 32]. Although annealing restores the β-phase in both radiation-induced sublayers [33], the recovery for fluences around this threshold is only partial, i.e., within the γ-layer region, while in the amorphous region, the signal remains at the random level in the aligned spectra (see Table 1 and Figure S1 in Supplementary Materials (SM)). Following the work of Azarov et al. [30], misorientation of the regrown β-Ga$_2$O$_3$ layer could be suspected. However, the DB-VEPAS and VEPALS results presented in Figure 4 demonstrate that it represents only one of several possible scenarios for this phenomenon.

It is worth also adding that due to these mentioned complications connected with the radiation-induced β → γ phase transition (above $1\times10^{14}$ ions/cm$^2$) after implantation, as well as the potential misorientation that may occur when the β phase is restored after annealing, the estimation of lattice site locations of different RE in the β-Ga$_2$O$_3$ matrix based on RBS/c spectra presented in Figures 1b and 3 (Figue S1 in SM) is impossible. In the present case, to ensure the optimal optical efficiency [14], the implantation energy was selected to modify only the top

~60 nm of the layer, making alignment along other possible crystallographic axes very difficult within such thin sublayers, as successfully demonstrated by Bektas et al. [28]. Additionally, it cannot be conclusive also for lower fluences (below $1\times10^{14}$ ions/cm$^2$), where this transformation does not occur, because the RBS signal from RE atoms is below the detection limit (see Figure 1b).

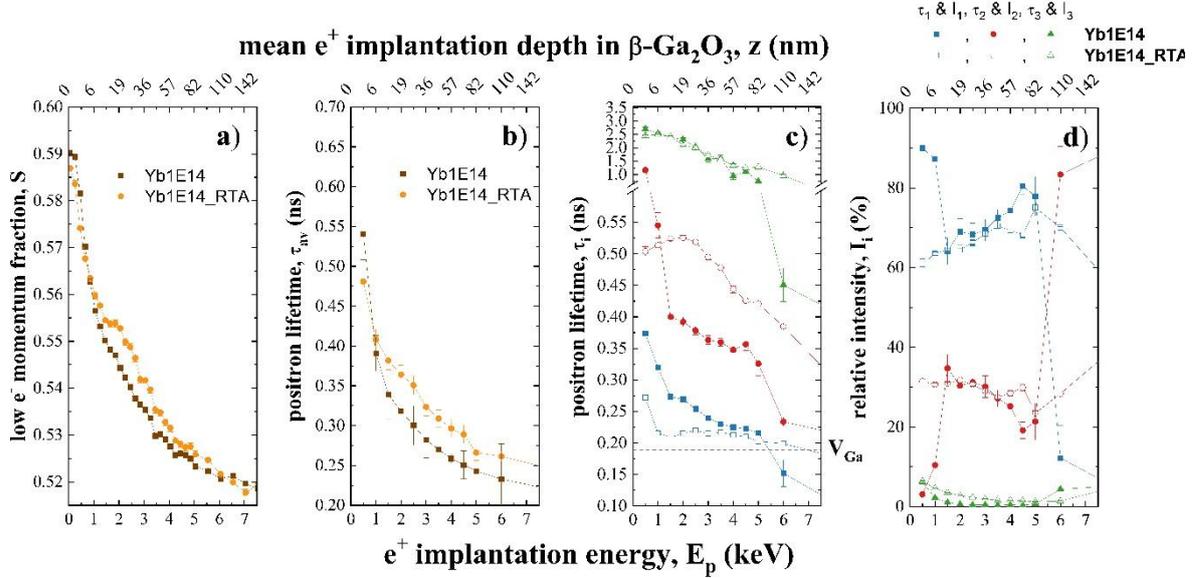

*Figure 4*. Annihilation line parameter S (low electron momentum fraction) (a), average positron lifetime $\tau_{av}$ (b), positron lifetime components (c), and their relative intensities (d) as a function of positron implantation energy $E_P$ and mean positron implantation depth z.

The DB-VEPAS measurement results obtained for Yb-implanted β-Ga$_2$O$_3$ and subsequently annealed samples are represented by the depth profiles of the annihilation line parameter S (low electron momentum fraction) and are presented in Figure 4a. For details, refer to the experimental and SM sections. The S-parameter reveals the contributions of valence electrons to the annihilation process, and scales mostly with defect concentration of positron traps such as neutral and negatively charged vacancies, vacancy clusters, dislocations, and larger free-volume defects, but it also depends on the defect size to some extent [36]. Figures 4b, c, and d show the positron lifetime (VEPALS) results, which reflect the local electron density and offer additional insight into the size and nature of these defects. Both DB-VEPAS and VEPALS measurements were performed as a function of positron implantation energy $E_P$, enabling depth profiling of the defects in β-Ga$_2$O$_3$. For details of the depth estimation, lifetime, and other spectra parameters calculations, please see Methods in the Supplementary Materials. In general, DB-VEPALS and VEPALS measurements demonstrate that annealing does not lead to the complete removal of radiation-induced defects in β-Ga$_2$O$_3$. Instead, it promotes their rearrangement and transformation.

The S-parameter (Figure 4a) and average positron lifetime τ$_{av}$ (Figure 4b) indicate similar trends across the sample depth, both slightly increasing after annealing in the sub-surface region of ~100nm. The increase is direct evidence of the overall rise in the average defect size. As illustrated in Figure 4c, the analysis of PALS spectra revealed the presence of three distinct lifetime components, noted τ$_1$, τ$_2$, and τ$_3$, which translates to the existence of three types of defects. The dashed line in Figure 4c presents the theoretical positron lifetime calculation of gallium monovacancy (V$_{Ga}$) [37], which equals 189 ps. The τ$_1$ component, which has the shortest lifetime, is longer than the lifetime expected for V$_{Ga}$ and can be attributed to a divacancy or a slightly larger vacancy cluster. In contrast, τ$_2$ corresponds to a much larger vacancy agglomerate. The τ$_3$ component, characterized by lifetimes exceeding 500 picoseconds (ps), can be attributed to void-like defects [38,39]. However, their concentration remains low, approaching the detection limit. Following the annealing process, significant changes can be observed. Both parameters S (Figure 4a) and average lifetime (Figure 4b) increase after annealing, but focusing on Figure 4c can be noticed that in fact the τ$_2$ lifetime increases after annealing, while τ$_1$ exhibits a slight decrease. This can be interpreted as a significant growth in size of larger clusters (τ$_2$) and the size reduction of smaller clusters (τ$_1$). Interestingly, the relative intensities remain similar across depth after annealing, which could be misinterpreted as unchanged defect density. However, the positron trapping yield increases for larger defects [36], hence the increase of τ$_2$ is a fingerprint of defect agglomeration and at the same time their density reduction. The opposite is expected for τ$_1$ related defects, as their size reduces after annealing, thus their concentration should increase. The absolute defect concentration is difficult to predict as no bulk positron lifetime is observed, hence positron saturation trapping is expected, which is typical for defect densities larger than 10$^{19}$ /cm$^2$ [40]. Additionally, some fluctuations in the intensity of τ$_1$ and τ$_2$ lifetimes (I$_1$ and I$_2$) can be observed after implantation in the near-surface region (0–10 nm), suggesting local variations in defect density and concentration. After annealing, these fluctuations vanish, indicating that the defect microstructure becomes more uniform. Finally, a slight increase can also be noticed in the intensity of the long-lived τ$_3$ (void-related) component after annealing, though its overall contribution still remains negligible. Following the information presented in Figure S2 in the SM, where the VEPALS results obtained for annealed samples implanted with different RE are compared, it can be concluded that the present behavior is common for all RE and the defect microstructure is similar for all of them, too. The structural results presented thus far do not answer the question of the nature of the potential clusters, and further studies are necessary.

## 3.2 Optical properties of β-Ga₂O₃:RE systems

In order to gain a comprehensive understanding of the optical activity of intrinsic defects and non-intentional impurities, as well as selected rare-earth dopants in β-Ga₂O₃, we analyzed the PL and PLE responses of both the as-grown material (unimplanted) and samples implanted with selected RE ions.

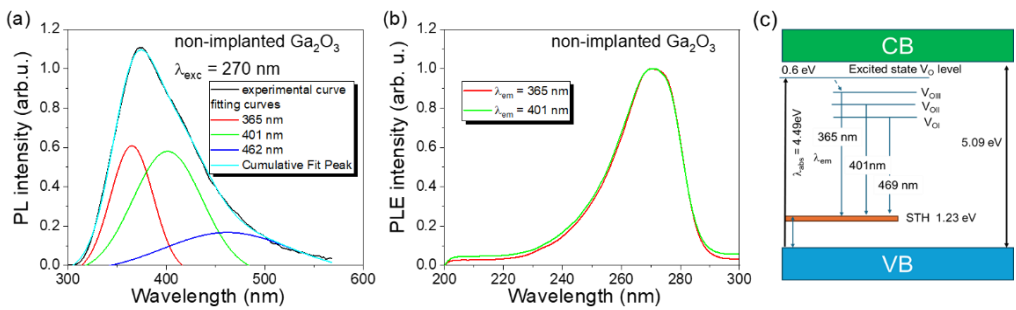

*Figure 5. PL and PLE of an unimplanted β-Ga₂O₃ crystal. (a) PL from oxygen vacancies excited with 270 nm, (b) normalized PLE of 365 nm and 401 nm emission lines presented in (a); (c) the schematic excitation mechanism of oxygen vacancies in gallium oxide; $V_{OI}$, $V_{OII}$, and $V_{OIII}$ are oxygen vacancies with different structures, and the STH is the self-trapped hole level, typically located about 1.23 eV above the valence band maximum.*

Figure 5 presents the characteristic PL emission (a) and normalized PLE (b) of the unimplanted (as-grown) β-Ga₂O₃ crystal oriented along the (-201) crystal axis. As can be seen, unimplanted gallium oxide exhibits strong emission in the UV–visible spectral range. Typically, the UV–visible emission from β-Ga₂O₃ is attributed to the radiative recombination from oxygen vacancies at different ionization states [9]. The obtained PL spectrum can be fitted with three Gaussian functions corresponding to the emissions at about 365 nm, 401 nm, and 462 nm. To monitor the two strongest peaks, the photoluminescence excitation measurements were performed. As it can be seen in Figure 5b, the normalized PLE spectra emissions at both 365 nm and 401 nm exhibit nearly identical excitation profiles, with a maximum absorption band located at approximately 275 nm (4.49 eV). The obtained value is significantly lower than the typically reported direct band gap of β-Ga₂O₃ (4.8 – 5.09 eV), which suggests that the excitation mechanism of oxygen vacancies in gallium oxide is more complex than a simple band-to-band transition. Dong et al. [9] investigated the structural configurations of oxygen vacancies ($V_O$) in β-$Ga_2O_3$ crystal and their influence on the optical properties of the matrix. According to these results, the basic configuration of $V_O$ strongly depends on the β-$Ga_2O_3$ symmetry configuration, and several distinct $V_O$ configurations can be distinguished in such a matrix: two with 3-fold symmetry ($V_{OI}$ and $V_{OII}$) and one with 4-fold symmetry ($V_{OIII}$). The energy levels of these vacancies are stable when they are neutral and are located above the valence band (VB) maximum at approximately 2.63, 3.10, and 3.40 eV for $V_{OI}$, $V_{OII}$, and $V_{OIII}$, respectively. As was already mentioned, deconvolution of the PL spectrum presented in Figure 5a results in three emissions at about 365 nm (3.38 eV), 401 nm (3.08 eV), and 462 nm (2.66 eV), that overlap very well with the calculated values presented by Dong et al. for neutral oxygen vacancies if the self-trapped hole (STH) level is considered. Figure 5c shows a schematic illustration of the possible excitation and emission mechanism of oxygen vacancies in gallium oxide. The radiative transition between the excited state and valence band (VB) has a low probability due to the formation of the STH level in gallium oxide. The STH is located about 1.23 eV above the VB and is partially responsible for the challenges in the p-type doping of gallium oxide [41]. At room temperature (RT), holes are localized at the STH level and do not contribute to the p-type conduction mechanism, but they can be active in the radiative recombination process. We postulate that the oxygen vacancies possess a more complex structure, with an excited state located at about 0.6 eV below the conduction band (CB) minimum. First, carriers are excited from VB to this high-energy $V_O^*$ excited state ($α_{exc}$ = 4.49 eV). Subsequently, through phonon-assisted non-radiative relaxation, they are transferred to lower-lying $V_O$ excited levels, which finally de-excite radiatively to the STH level. Deep-level transient spectroscopy (DLTS) studies revealed the existence of a deep electron trap level, E1, located at about 0.55 – 0.6 eV below the CB minimum [42, 43]. The authors suggested that the E1 level is due to some metallic impurities. On the other hand, the moderate crystalline quality typically contains multiple donor-like defects that are responsible for unintentional n-type doping. The free-carrier absorption associated with these donors can reduce the effective band gap by approximately 100–500 meV. Since oxygen vacancies are mainly responsible for the n-type conductivity, the attribution of the $E_1$ level to $V_O$-related defects appears reasonable.

Building on this understanding of the origin and excitation pathways of the intrinsic defect-related luminescence in β-$Ga_2O_3$, next, the optical response of β-$Ga_2O_3$ doped with different rare-earth ions was investigated using PL and PLE spectroscopy at RT. Figure 6 shows the photoluminescence emission (a) and excitation (b) spectra obtained from samples implanted with Dy, Er, and Yb ions to a fluence of $1\times10^{15}$ /cm$^2$ and subsequently annealed at 800 °C for 10 min in an oxygen atmosphere. As shown in Figure S3, for the β-Ga$_2$O$_3$ systems implanted with this fluence, the luminescence efficiency of Yb$^{3+}$ is the highest, despite the considerable lattice disorder. Above $1\times10^{15}$ /cm$^2$, the luminescence starts to be quenched, most probably due to the concentration quenching effect [44].

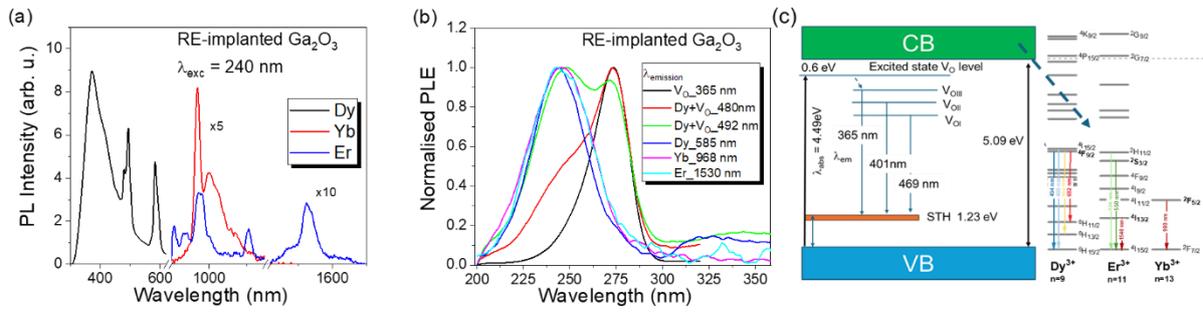

*Figure 6. Photoluminescence emission (a) and excitation (b) spectra of β-Ga$_2$O$_3$ systems doped with different RE. (c) schematic excitation mechanism of RE in β-Ga$_2$O$_3$.*

As shown in Figure 6a, the Dy-implanted β-Ga$_2$O$_3$ system exhibits the broad emission in the UV–visible spectral range of with the dominant band centered at ~365 nm, attributed to V$_{OIII}$-related centers. In addition to the defect-related emission, the sample exhibits several sharp Dy$^{3+}$ f–f transitions that are observed at approximately 480 nm, 492 nm, and 585 nm, corresponding to the radiative transitions between $^4I_{15/2}$ and $^4F_{9/2}$ excited states and $^6H_{13/2}$ and $^6H_{15/2}$ ground state multiplets. The main PL emission from Er$^{3+}$ is detected in the infrared region at ~1540 nm due to the optical transition between $^4I_{13/2}$ and the ground state $^4I_{15/2}$, while the characteristic PL emission from Yb$^{3+}$ in β-Ga$_2$O$_3$ is observed at ~980 nm and is associated with the $^2F_{5/2}\rightarrow{}^2F_{7/2}$ transition.

In order to investigate the excitation mechanism of RE$^{3+}$ ions in β-Ga$_2$O$_3$, PLE spectra were recorded by monitoring the corresponding emissions as a function of excitation wavelength in the 200–350 nm range. As illustrated in Figure 2b, two distinct excitation bands are observed: one at about 240 nm and the second at about 275 nm. The peak at ~275 nm appears when PL emission is fixed at 365 nm, 480 nm, and 492 nm—i.e., in the spectral region in which the PL spectra correspond to the oxygen-vacancies related emission. The pure Dy$^{3+}$ emission at 585 nm exhibits a maximum excitation at ~240 nm, which is also the dominant excitation wavelength for the characteristic Er$^{3+}$ and Yb$^{3+}$ infrared emissions. A similar excitation behavior was observed from Pr$^{3+}$ doped samples by Zanoni et al. and from Tb$^{3+}$ by Tokida and Adachi [5, 17]. It suggests that all trivalent RE ions embedded into β-Ga$_2$O$_3$ have the same excitation mechanism, and it is not necessarily the non-radiative relaxation from the 4f5d state to the 4f levels proposed recently [5]. In fact, the 5d level of RE has a similar location positioned near the conduction band (CB) minimum of wide-band-gap hosts, but their exact energies are not the same as demonstrated by systematic studies of 4f–5d transitions in fluorides, halogenides, and chalcogenides [45-46]. Rodnyi et al. [45] have presented the location of 5d energy levels of different RE$^{3+/2+}$ in BaF$_2$ and CdF$_2$ and clearly shown that 5d energy levels change with the number of 4f electrons. In turn, Dorenbos [46] has summarized the 4f↔5d transitions of the trivalent lanthanides in halogenides and chalcogenides, and also pointed out that for different rare earths, the position of the 5d level varies significantly between different matrices. In this light, to interpret the PLE results presented here, we postulate that the RE$^{3+}$ ions in β-Ga$_2$O$_3$ are directly excited via the host conduction band. First, electrons are excited to the CB of β-Ga$_2$O$_3$ ($\lambda_{exc} \approx 240$ nm). Then, they relax via non-radiative phonon transfer to the 4f excited states of the RE$^{3+}$ ions. Finally, they radiatively de-excite to the respective 4f ground states of the RE$^{3+}$ ions with photon emission. The schematic of the RE$^{3+}$ excitation mechanism in β-Ga$_2$O$_3$ is shown in Figure 6c. The PLE peak at approximately 240 nm (5.1 eV) matches well with the band gap of β-Ga$_2$O$_3$. Meanwhile, the second PLE peak at approximately 280 nm (0.6 eV below the CB minimum) is related to the excited V$_O$-related state. It is noteworthy that an analogous excitation mechanism of RE$^{3+}$ ions has been observed in a ZnO matrix, where the PLE peak overlaps with the band gap energy of ZnO [47].

## 4 Conclusions

In this work, we have conducted a comprehensive study of the structural and optical properties of β-Ga$_2$O$_3$ single crystals implanted with different fluences of Dy, Er, and Yb ions and subsequently annealed in oxygen at 800 °C for 10min., using a wide range of complementary characterization techniques. Based on RBS/c analysis, we have shown that the implantation-induced disorder, associated β→γ phase transition, and subsequent structural recovery after annealing proceed in a largely similar manner for all investigated RE ions. Using PAS, we have further demonstrated that annealing does not eliminate radiation-induced defects but likely increases the concentration of small vacancy clusters. Instead, we find that part of defects rearrange and evolve into larger complexes. The phenomenon also seems to be common among the studied RE dopants, since they defect landscape is very similar to each other.

The optical measurements confirmed that unimplanted β-Ga$_2$O$_3$ exhibits strong UV–visible emission governed by oxygen-vacancy–related centers. We have shown that the introduction of RE ions adds characteristic sharp lines, originating from the transition of a given RE$^{3+}$. Based on PLE results, we postulate that all RE$^{3+}$ ions in β-Ga$_2$O$_3$ are excited via the host conduction band, followed by non-radiative relaxation to 4f excited states and radiative decay to their respective ground states. Furthermore, by analyzing the fluence dependence of Yb$^{3+}$ emission, we have identified the onset of concentration quenching and discovered that Yb$^{3+}$ maintains the highest emission efficiency even in the presence of substantial lattice disorder.

Overall, we have provided new insights into defect evolution in ion-implanted β-Ga$_2$O$_3$ and the excitation mechanisms of RE$^{3+}$ ions in this host material. We believe that these findings offer valuable guidelines for optimizing β-Ga$_2$O$_3$:RE systems for future optoelectronic applications.

**Acknowledgements**

The research was carried out as part of the NCN project UMO-2022/45/B/ST5/02810. Parts of this research were conducted at the IBC, ELBE, and the Materials Department of the Helmholtz-Zentrum Dresden – Rossendorf e. V., a member of the Helmholtz Association. We would like to thank the facility staff for their assistance. The DBS experimental work was also partially supported by the Initiative and Networking Fund of the Helmholtz Association (FKZ VH-VI-442 Memriox), and the Helmholtz Energy Materials Characterization Platform (03ET7015).

# Supplementary Materials

Structural and Optical Characteristics of β-Ga$_2$O$_3$ Implanted with Rare Earth Ions

*Renata Ratajczak\*, Joanna Matulewicz, Slawomir Prucnal, Maciej O. Liedke, Cyprian Mieszczynski, Przemyslaw Jozwik, Ulrich Kentsch, Rene Heller, Eric Hirschmann, Andreas Wagner, Wojciech Wozniak, Frederico Garrido and Elzbieta Guziewicz*

# Figures

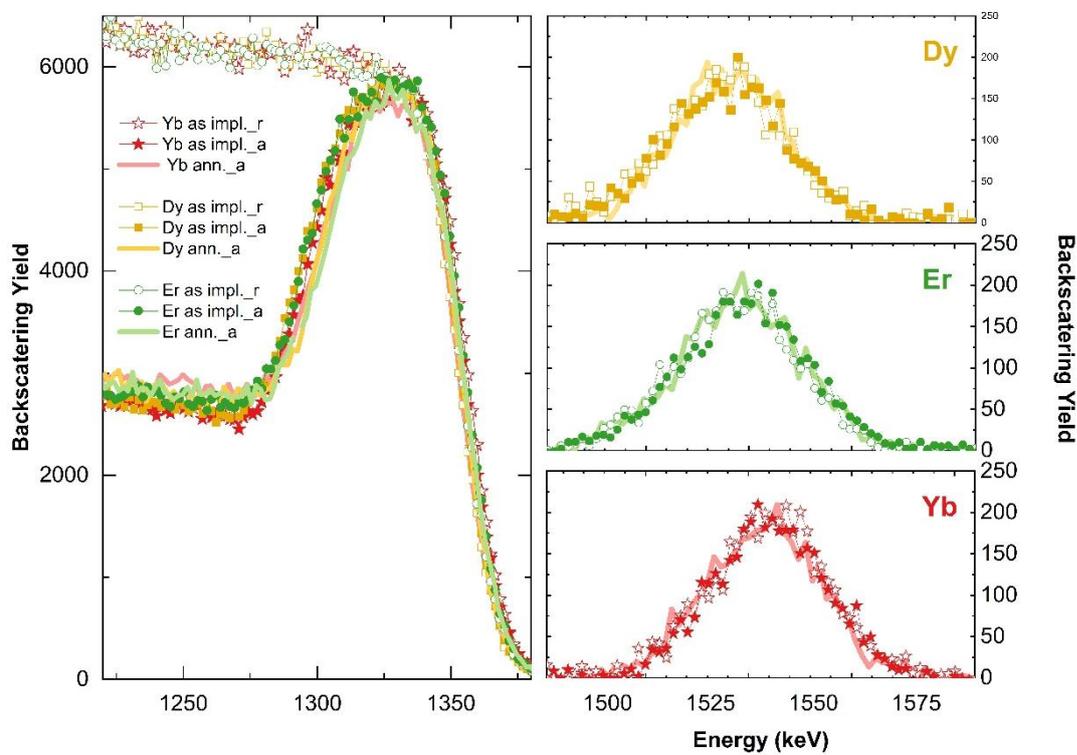

Figure S1 Comparison of random (r) and aligned (a) RBS spectra obtained for (-201)-oriented β-Ga$_2$O$_3$ single crystals implanted with Dy, Er, and Yb ions to a fluence of 1x10$^{15}$ /cm$^2$ prior to and after annealing in oxygen atmosphere at 800°C for 10min.

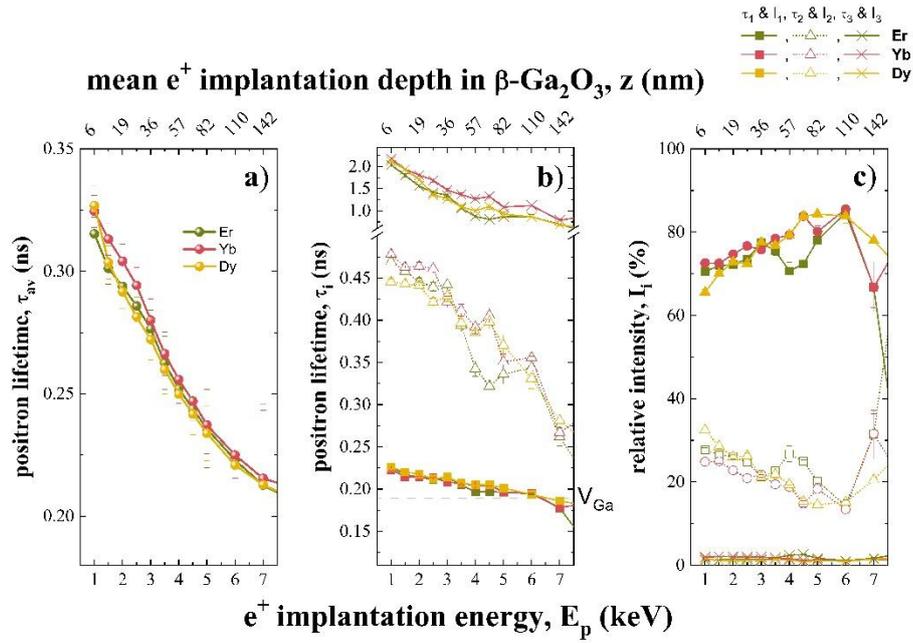

Figure S2 VEPALS results for post-annealed β-Ga$_2$O$_3$ implanted with Dy, Er, and Yb to a fluence of 1x10$^{15}$ /cm$^2$. Panels (a) show the average positron lifetime $\tau_{av}$, positron lifetime components (b), and their relative intensities (c) as a function of positron implantation energy $E_p$ and mean positron implantation depth <z>. The dashed line in (b) represents the theoretically calculated lifetime for Ga-vacancy (V$_{Ga}$) in β-Ga$_2$O$_3$ [1]

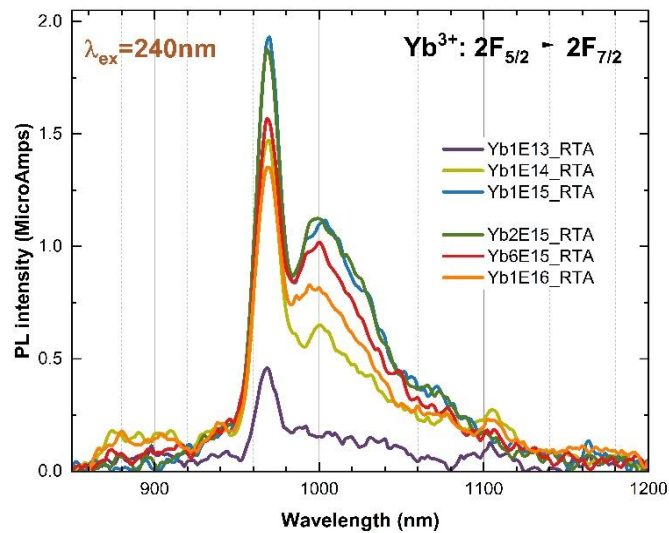

Figure S3 PL spectra collected at RT in the IR spectral range for (-201) oriented β-Ga$_2$O$_3$ single crystals implanted with different fluences of Yb ions and annealing in oxygen

atmosphere at 800°C for 10min. The figure shows fluence-dependent studies of $Yb^{3+}$ into $\beta$-$Ga_2O_3$ matrix.

## Methods
Doppler broadening variable energy positron annihilation spectroscopy (DB-VEPAS)

Doppler broadening variable energy positron annihilation spectroscopy (DB-VEPAS) measurements have been conducted at the apparatus for in-situ defect analysis (AIDA) [2] of the slow positron beamline (SPONSOR) [3]. Positrons have been accelerated and monoenergetically implanted into samples in the range between $E_p$ = 0.05-35keV, which allows for depth profiling. A mean positron implantation depth was approximated using a simple material density ($\rho$) dependent formula [4]:

$$\langle z \rangle \text{ [nm]} = \frac{36}{\rho \text{ [g·cm}^{-3}]} E_p^{1.62} \text{ [keV]}.$$

‹z› approximates the depth and cannot be treated as an absolute measure because it does not account for positron diffusion. The best estimation it gives for materials with large defect concentration, hence low positron diffusion length. Implanted into a solid, positrons lose their kinetic energy due to thermalization and, after short diffusion, annihilate in delocalized lattice sites or localize in vacancy-like defects and interfaces, emitting usually two anti-collinear 511keV gamma photons once they meet electrons. Since at the annihilation site thermalized positrons have very small momentum compared to the electrons, a *broadening of the 511 keV line* is observed mostly due to the momentum of the electrons, which is measured with one or two high-purity Ge detectors (energy resolution of 1.09 ± 0.01 keV or 0.78 ± 0.02 keV at 511 keV for single and double detector configuration). This broadening is characterized by two distinct parameters *S* and *W*, defined as a fraction of the annihilation line in the middle (511±0.93 keV) and outer regions (504.04-508.68 keV and 513.32-517.96 keV), respectively. The total area below the curve, which is utilized for normalization of both parameters, is 511±16.24 keV. The S-parameter is a fraction of positrons annihilating with low-momentum valence electrons and represents vacancy-type defects and their concentration. The W-parameter approximates the overlap of the positron wavefunction with high-momentum core electrons. Plotting calculated *S* as a function of positron implantation energy, $S(E_p)$, provides depth-dependent information, whereas *S-W* plots are used to examine the atomic surroundings of the defect site and its size (type). [5,6]

**Variable energy positron annihilation lifetime spectroscopy (VEPALS)**

Variable energy positron annihilation lifetime spectroscopy (VEPALS) measurements were conducted at the Mono-energetic Positron Source (MePS) beamline at HZDR, Germany [7]. A CeBr$_3$ scintillator detector coupled to a Hamamatsu R13089-100 photomultiplier tube (PMT) was utilized for gamma photon detection. The signals were processed by the Teledyne SP Devices ADQ14DC-2X digitizer (14-bit vertical resolution and 2GS/s horizontal resolution) [8]. The overall time resolution of the measurement system is ≈0.250 ns, and all spectra contained at least $1 \cdot 10^7$ counts. A typical lifetime spectrum *N(t)*, the absolute value of the time derivative of the positron decay spectrum, is described by

$$N(t) = R(t) * \sum_{i=1}^{k+1} \frac{I_i}{\tau_i} e^{\frac{-t}{\tau_i}} + \text{Background},$$

where *k* is the number of different defect types contributing to the positron trapping, which are related to *k* + 1 components in the spectra with the individual lifetimes $\tau_i$ and intensities $I_i$ ($\Sigma I_i=1$) [9]. The instrument resolution function *R*(*t*) is a sum of two Gaussian functions with distinct intensities and relative shifts, both depending on the positron implantation energy, E$_p$. It was determined by the measurement and analysis of a reference sample, i.e., Yttria-stabilized zirconia (YSZ), which exhibited a single well-known lifetime component. The background was negligible, hence fixed to zero.

All the spectra were deconvoluted using a non-linear least-squares fitting method, minimized by the Levenberg-Marquardt algorithm, employed within the fitting software package PALSfit [10] into 2 major lifetime components, which directly evidence localized annihilation at 2 different defect types (sizes; $\tau_1$ and $\tau_2$). 2 minor lifetime components ($\tau_3$ and $\tau_4$) were necessary in order to obtain a fit with high goodness. They represent positron annihilation at micropores (voids) and ortho-Positronium residual signals or spectra contaminations, respectively. Their relative intensities scale typically with the concentration of each defect type but are affected by positron trapping yields. In general, positron lifetime increases with defect size and open volume size. The positron lifetime and its intensity have been probed as a function of positron implantation energy $E_p$, which was recalculated to the mean implantation depth ‹z›. The average positron lifetime $\tau_{av}$ is defined as $\tau_{av} = \sum_i \tau_i \cdot I_i$.